\documentclass[footinbib,twocolumn,showpacs,amsmath,amstex,amssymb,mathfonts,superscriptaddress,prl]{revtex4}
\usepackage{graphicx}
\usepackage{color}
\usepackage{bm}

\usepackage{graphicx}
\usepackage{color}
\usepackage{bm}
\usepackage{amsmath}
\usepackage{amssymb}
\usepackage{amsthm}
\usepackage{amsfonts}
\usepackage{multirow}
\usepackage{makecell}
\usepackage{float}
\usepackage{comment}
\usepackage{enumitem}
\usepackage{verbatim}

\usepackage{bbm}

\begin{document}

\title{Fractional quantum Hall effect from frustration-free Hamiltonians}
\author{Bo Yang} 
\affiliation{Division of Physics and Applied Physics, Nanyang Technological University, Singapore 637371.}
\affiliation{Institute of High Performance Computing, A*STAR, Singapore, 138632.}

\date{\today}
\begin{abstract}
We show that there is an emergent lattice description for the continuous fractional quantum Hall (FQH) systems, with a generalised set of few-body coherent states. In particular, model Hamiltonians of the FQH effect are equivalent to the real space von Neumann lattice of local projection operators imposed on a continuous system in the thermodynamic limit. It can be analytically derived that tuning local one-body potentials in such lattices amounts to the tuning of individual two or few-body pseudopotentials. For some cases, we can realise pure few-body pseudopotentials important for stabilising exotic non-Abelian topological phases. This new approach can thus potentially lead to experimental realisation of coveted non-Abelian quantum fluids including the Moore-Read state and the Fibonacci state. The reformulation of the FQHE as a sum of local projections opens up new path for rigorously proving the incompressibility of microscopic Hamiltonians in the thermodynamic limit.
\end{abstract}

\maketitle 

The fractional quantum Hall (FQH) systems are promising candidates for hosting anyonic and non-Abelian excitations that are topologically protected with strong interactions between electrons\cite{mr,rr}. With these exotic properties, the relevant two-dimensional electron gas systems are robust for storing and processing quantum information, serving as potential physical platforms for topological quantum computers\cite{sdsarma}. Experimentally, however, there has been no conclusive evidence for the non-Abelian statistics, even for the simplest non-Abelian FQH states: the Moore-Read (MR) state at half filling\cite{willett1,willett2,lin1,willett3,lin2,west}. For universal quantum computation, the theoretically proposed Fibonacci FQH state\cite{barkeshli,zhuwei} at filling factor $\nu=3/5$ is needed. While the MR state is constructed as the ground state of the model Hamiltonian with a three-body interaction, the model Hamiltonian for the Fibonacci state requires a much more complex four-body interaction\cite{rr}. Many other interesting FQH states have been proposed theoretically with model Hamiltonians in the form of pseudopotentials, but the rich topological structures of these states are largely beyond the reach of experiments at the current stage. 

One of the main challenges for exotic non-Abelian FQH states is that the model Hamiltonians are highly artificial. While in realistic systems the electron-electron interaction is derived from the two-body Coulomb interaction, exact non-Abelian FQH states typically require three- or more body interactions\cite{simonproj}. Numerical studies for small systems seem to indicate that it is possible for two-body realistic interactions to be adiabatically connected to some simple artificial model Hamiltonians based on wavefunction overlaps, and most works focused on the Moore-Read state\cite{peterson,peterson1,wanghao}. It is, however, difficult to deduce the topological properties of realistic systems in the thermodynamic limit. This is especially true for the non-Abelian states, in which the non-Abelian statistics are determined by degeneracies of elementary excitations. Thus the ground state incompressibility gap, as well as the low-lying excitation bandwidth, need to be carefully tuned in experiments\cite{toke,yang1}. Unfortunately, experimental tuning of the electron-electron interaction is highly constrained. There are also limited theoretical guidance on the optimal Coulomb based Hamiltonians that can mimic three or more body interactions\cite{rezayi,yang2}.

Theoretically, there are also fundamental and unproven questions on the incompressibility of model Hamiltonians (or realistic Hamiltonians close to them)\cite{rougerie}. Despite overwhelming numerical evidence, there is no rigorous proof that the model Hamiltonians are gapped in the thermodynamic limit, even for the simplest Laughlin states. It is also conjectured that certain model Hamiltonians (e.g. for the Gaffnian state) are gapless\cite{gaffnian,read2}, but whether the gap closes in the same or different quantum sector as the ground state has important implications for the topological nature of the FQH phase\cite{yang1}. For example, the Gaffnian and the Jain ground state state at $\nu=2/5$ could be topologically equivalent if all translationally invariant excitations are gapped. For strongly correlated systems, there are rigorous proofs in spin systems when the model Hamiltonians are the sum of real-space projections (e.g. the AKLT model\cite{aklt1,aklt2,aklt3,aklt4}). In contrast, FQH model Hamiltonians are projections of particle cluster angular momenta in the continuum limit, so the same methodologies do not seem to apply. 

In this Letter, we show the equivalence of the von Neumann lattice (vNL) of local projections in real space and conventional model FQH Hamiltonians. Explicit derivations are presented for the Read-Rezayi series, showing FQH states (including non-Abelian, exotic ones such as the Moore-Read state and the Fibonacci state) can be realised with a vNL of properly tuned local potentials. The construction of the local projections is reminiscent of the classical local exclusion conditions proposed in Ref.\cite{yang3}, but the lattice of such projections is a well-defined quantum Hamiltonian in the continuum that can in principle be realised experimentally. There are interesting analogies between this new form of FQH Hamiltonians and the spin Hamiltonian for the AKLT model. The latter can be shown rigorously to have a gap due to the unique properties of the projection operators. Thus the equivalence we establish here provides new routes for potentially proving the incompressibility of FQH Hamiltonians in the thermodynamic limit.

{\it The projection Hamiltonians --}
It has been established numerically\cite{yang3} that the model ground states of many FQH phases can be uniquely determined by two constraints: translational/rotational invariance and the classical reduced density matrix constraint. The latter is denoted with a triplet of non-negative integers $\hat c=\{n,n_e,n_h\}$. Physically, it dictates for any small droplet containing $n$ fluxes (thus with an area of $2n\pi l_B^2$, $l_B$ being the magnetic length), a measurement in this droplet can never detect more than $n_e$ number of electrons or $n_h$ number of holes (unoccupied orbitals in a single LL). For example, $\hat c_l=\{2,1,2\}$ dictates that no more than one electron and no more than two holes can be detected in any circular droplet containing two fluxes in the quantum fluid. It gives the Laughlin state at filling factor $\nu=1/3$ and topological orbital shift $S_h=-2$. Another example is $\hat c_f=\{4,3,4\}$, giving the Fibonacci state at $\nu=3/5$ and $S_h=-2$. 

These classical constraints do not correspond exactly to any local Hamiltonians, and cannot be exactly implemented in real space. We can, however, define the local projection operator as follows:
\begin{eqnarray}\label{h0}
\mathcal H_0=\sum_i|\psi_i\rangle\langle\psi_i|
\end{eqnarray}
where each state $|\psi_i\rangle$ consists of $n$ orbitals of the symmetric gauge around the origin in a single LL, and the summation is over all such states that \emph{do not} satisfy $\hat c=\{n,n_e,n_h\}$. Diagonalisation of Eq.(\ref{h0}) in the sub-Hilbert space of translationally invariant states will lead to the ground state of the corresponding FQH state. Within the full Hilbert space, however, we cannot obtain the FQH ground state because Eq.(\ref{h0}) is not translationally invariant.

To construct a translationally invariant analog of Eq.(\ref{h0}), we briefly review the well-known vNL formalism in a single LL, and introduce the notations\cite{dana,tanaka,von}. Using the guiding center coordinates $\hat R^x, \hat R^y$ with the commutation relation $[\hat R^x,\hat R^y]=-il_B^2$, these operators only have matrix elements within a single LL. We can construct the ladder operators $\hat b=\left(\hat R^x-i\hat R^y\right)/\sqrt 2l_B$ and $[\hat b,\hat b^\dagger]=1$. Denoting the single particle state centered at the origin as $|n\rangle=1/\sqrt{n!}\left(\hat b^\dagger\right)^n|0\rangle$, with $\hat b|0\rangle=0$. Thus $|0\rangle$ is the coherent state at the origin.

Given the commutation relation of $\hat R^x, \hat R^y$, the magnetic translation operator is given by $\hat T_{\vec X}=\prod_ie^{iX_a\hat R_i^a}$, where Einstein's summation convention is adopted and the subscript $i$ runs over all electrons. The state $|\vec X\rangle=\hat T_{\vec X}|0\rangle$ is thus the coherent state centered at $r^a=l_B^2\epsilon^{ab}X_b$ in the real space. These states are not orthogonal, but form a complete basis with $\vec X$ as a continuous variable. This basis is obviously over-complete, since the total number of linearly independent states in a single LL is $\mathcal A/\left(2\pi l_B^2\right)$, where $\mathcal A$ is the area of the sample. A minimal complete basis of the coherent states can be formed from a square vNL with $r^x=\sqrt{2\pi}l_Bp, r^y=\sqrt{2\pi}l_Bq$, where $p,q$ are integers. Rewriting $|\vec X\rangle=|p,q\rangle$, we thus have $\sum_{p,q=-\infty}^{\infty}|p,q\rangle\langle p,q|=\mathcal I$. This relationship also works if we replace $|0\rangle$ with $|n\rangle$ in the above analysis.

The generalisation of Eq.(\ref{h0}) at a single site to the vNL is straightforward, since we can form the vNL where each site contains more than one electron. Denoting $|\psi(\vec 0)\rangle$ as a state of a circular droplet centered at the origin, we have $|\psi(\vec X)\rangle=\hat T_{\vec X}|\psi(\vec 0)\rangle$. The resulting state describes the circular droplet centered at $r^a=l_B^2\epsilon^{ab}X_b$. In the thermodynamic limit with $\mathcal A\rightarrow\infty$ or $l_B\rightarrow 0$, the vNL effective Hamiltonian can thus be constructed as follows:
\begin{eqnarray}\label{h}
\mathcal H=\int\frac{d^2r}{2\pi l_B^2}|\psi(\vec X)\rangle\langle\psi(\vec X)|
\end{eqnarray}
Due to the completeness of the vNL, Eq.(\ref{h}) is translationally invariant in a single LL. If $|\psi_i(\vec X)\rangle$ is a single electron state, Eq. (\ref{h}) is just the identity matrix. 

It turns out that Eq.(\ref{h}) becomes highly non-trivial when $|\psi_i(\vec X)\rangle$ contains two or more electrons. The simplest example is to look at $|\psi(\vec 0)\rangle=|0,1\rangle$, describing a droplet containing two fluxes, both occupied by electrons. Here we define $|s_1,s_2,\cdots s_{N_e}\rangle=\hat c_{s_1}^\dagger \hat c_{s_2}^\dagger\cdots \hat c_{s_{n_e}}^\dagger|0\rangle\sim\text{Asy}\left[\left(\hat b_1^\dagger\right)^{s_1}\left(\hat b_2^\dagger\right)^{s_2}\cdots \left(\hat b_{n_e}^\dagger\right)^{s_{n_e}}\right]|0\rangle$ as the $n_e-$ electron Slater determinant state, with $\text{Asy}$ denoting antisymmetrisation over electron indices (the subscript); $\hat c_i^\dagger$ is the second quantised electron creation operator. The two-body matrix elements of Eq.(\ref{h}) can be derived as follows:
\begin{eqnarray}\label{v}
&&\hat{\mathcal H}=\sum_{m,n,m',n'}V_{m,n}^{m',n'}\hat c^\dagger_{m'}\hat c^\dagger_{n'}\hat c_m\hat c_n\\
&&V_{m,n}^{m',n'}=\int\frac{d^2r}{2\pi l_B^2}\langle m',n'|\hat T_{\vec X}|0,1\rangle\langle 0,1|\hat T_{\vec X}|m,n\rangle\qquad
\end{eqnarray}

The integration over the entire two-dimensional plane can be carried out explicitly, and Eq.(\ref{v}) is equivalent to the well-known model Hamiltonian $\hat V_1^{\text{2bdy}}$ for the Laughlin state at $\nu=1/3$, given by the first Haldane pseudopotential. Thus the vNL of local potentials in the thermodynamic limit gives the exact spectrum of the Laughlin quantum Hall fluids. Such equivalence can be established analogously for the entire Read-Rezayi series. The three-body model Hamiltonian $\hat V_3^{\text{3bdy}}$ for the Moore-Read (MR) state can be obtained by taking $|\psi(\vec 0)\rangle=|0,1,2\rangle$. The four-body model Hamiltonian $\hat V_6^{\text{4bdy}}$ for the Fibonacci state can be obtained by taking $|\psi(\vec 0)\rangle=|0,1,2,3\rangle$.

We thus have a simple interpretation of the vNL of coherent states in a single LL, if we use these coherent states as projection operators in the form of Eq.(\ref{h}). Conventionally the coherent state is a droplet of one electron in one magnetic flux, and the completeness of the vNL leads to an identity matrix as the Hamiltonian, corresponding to the integer quantum Hall effect. If the coherent state is generalised to a droplet of $n$ electrons in $n$ magnetic fluxes, the corresponding vNL Hamiltonians are $n-$body model Hamiltonians of the Read-Rezayi series. Model Hamiltonians for many other FQH states (e.g. the Laughlin series, the Gaffnian and Haffnian states, etc) can be similarly shown to be equivalent to the vNL of projection operators from coherent states. This is because a coherent state can be a droplet containing any number of fluxes, with different arrangement of electrons in the droplet, in analogy to the LEC constraints introduced in Ref.\cite{yang3}.

{\it Experimental relevance --}
The derivation of model Hamiltonians from the vNL projectors not only reveals new perspectives on the physical nature of these Hamiltonians, it also naturally leads to a new way for the experimental realisation of exotic FQH fluids. If there are some mechanisms in mimicking Eq.(\ref{h0}) at a single location, e.g. with some local potential profile and Coulomb blockade effects, then a proper lattice pattern of such local mechanisms in principle can realise the effective projection Hamiltonians, and thus the topological phases in a robust manner. In particular, we can engineer a lattice of anti-dots of the size of a few magnetic fluxes, with tunable on-site potentials. Such a lattice will not break translational symmetry in a single LL, since the spacing between lattice points is on the order of $\sqrt{2\pi}l_B$. This could particularly be useful for few-body interactions, since LL mixing is not required. One can thus go to large magnetic field for large incompressibility gap.

In principle we can tune the relative strength of any few-body states on the anti-dots. Here we look at a simple but more restrictive effective local one-body anharmonic well, which is nevertheless an interesting example for tuning individual two- or more body pseudopotentials. We would like to emphasis that the set-up cannot be realised just by a simple one-body potential profile. It needs delicate Coulomb blockade-like effects that could be technically challenging. The effective potential well is given as follows:
\begin{eqnarray}\label{v0}
\hat V_0&=&\sum_{k=0}^n\lambda_k|k\rangle\langle k|\qquad
\end{eqnarray}
The upper limit $n$ in the summation gives the range of the local potential, covering an area of $\sim 2n\pi l_B^2$. Denoting a state containing $n_e$ number of electrons as $|k_1\cdots k_{n_e}\rangle$, with $k_i\le n$, a vNL of the local potential given by Eq.(\ref{v0}) leads to the effective Hamiltonian following the construction of Eq.(\ref{h}):
\begin{eqnarray}\label{potential}
&&\mathcal H=\int\frac{d^2r}{2\pi l_B^2}\sum\lambda_{k_1\cdots k_{n_e}}|k_1\cdots k_{n_e}\rangle\langle k_1\cdots k_{n_e}|\qquad\\
&&\lambda_{k_1\cdots k_{n_e}}=\lambda_{k_1}+\lambda_{k_2}+\cdots +\lambda_{k_{n_e}}
\end{eqnarray}
where all states not in the summation are presumably not present due to the blockade effect. The vNL of $n_e=1$ projections in Eq.(\ref{potential}) leads to a uniform background potential that can be ignored. Each vNL of $n_e$ electron projection leads to a linear combination of $n_e-$ body pseudopotential interactions. For example, $|0,1\rangle, |0,2\rangle$ both give $\hat V_1^{\text{2bdy}}$ only, $|0,3\rangle$ gives $\hat V_1^{\text{2bdy}}, \hat V_3^{\text{2bdy}}$; $|0,1,2\rangle, |0,1,3\rangle$ gives $\hat V_3^{\text{3bdy}}$, while $|0,1,2,3\rangle, |0,1,2,4\rangle$ gives $\hat V_6^{\text{4bdy}}$, and so on\cite{supp}. 

The coefficients of the linear combination of pseudopotentials for the vNL of each $|k_1\cdots k_{n_e}\rangle$ can be computed analytically. For example with $n=2$ (with $n$ from Eq.(\ref{v0}) giving the range of local potential), the corresponding vNL effective Hamiltonian is given by\cite{supp}:
\begin{eqnarray}\label{mr}
\hat{\mathcal H}&=&\int\frac{d^2r}{2\pi l_B^2}\lambda_{0,1}|0,1\rangle\langle 0,1|+\lambda_{0,2}|0,2\rangle\langle 0,2|\nonumber\\
&&+\lambda_{1,2}|1,2\rangle\langle 1,2|+\lambda_{0,1,2}|0,1,2\rangle\langle 0,1,2|\nonumber\\
&=&\left(2\lambda_{0,1,2}-\frac{3}{4}\lambda_{1,2}\right)\hat V^{\text{2bdy}}_1+\frac{3}{4}\lambda_{1,2}\hat V^{\text{2bdy}}_3\nonumber\\
&&+\frac{2}{3}\lambda_{0,1,2}\hat V_3^{\text{3bdy}}
\end{eqnarray} 
where we have ignored the uniform background. For $n=3$, we have the following instead\cite{supp}:
\begin{eqnarray}\label{fb}
\hat{\mathcal H}&=&\left(\frac{11}{4}\lambda_{0,1,2,3}-\lambda_{1,2,3}-\frac{3}{8}\lambda_{2,3}\right)\hat V_1^{\text{2bdy}}\nonumber\\
&+&\left(\frac{1}{4}\lambda_{0,1,2,3}+\lambda_{1,2,3}-\frac{1}{4}\lambda_{2,3}\right)\hat V_3^{\text{2bdy}}+\frac{5}{8}\lambda_{2,3}\hat V_5^{\text{2bdy}}\nonumber\\
&+&\left(\frac{7}{3}\lambda_{0,1,2,3}-\frac{8}{27}\lambda_{1,2,3}-\frac{2}{3}\lambda_{2,3}\right)\hat V_3^{\text{3bdy}}\nonumber\\
&+&\left(\frac{2}{3}\lambda_{0,1,2,3}-\frac{4}{9}\lambda_{1,2,3}+\frac{2}{3}\lambda_{2,3}\right)\hat V_5^{\text{3bdy}}\nonumber\\
&+&\frac{20}{27}\lambda_{1,2,3}\hat V_6^{\text{3bdy}}+\lambda_{0,1,2,3}\hat V_6^{\text{4bdy}}
\end{eqnarray} 

\begin{figure}[htb]
\includegraphics[width=6cm]{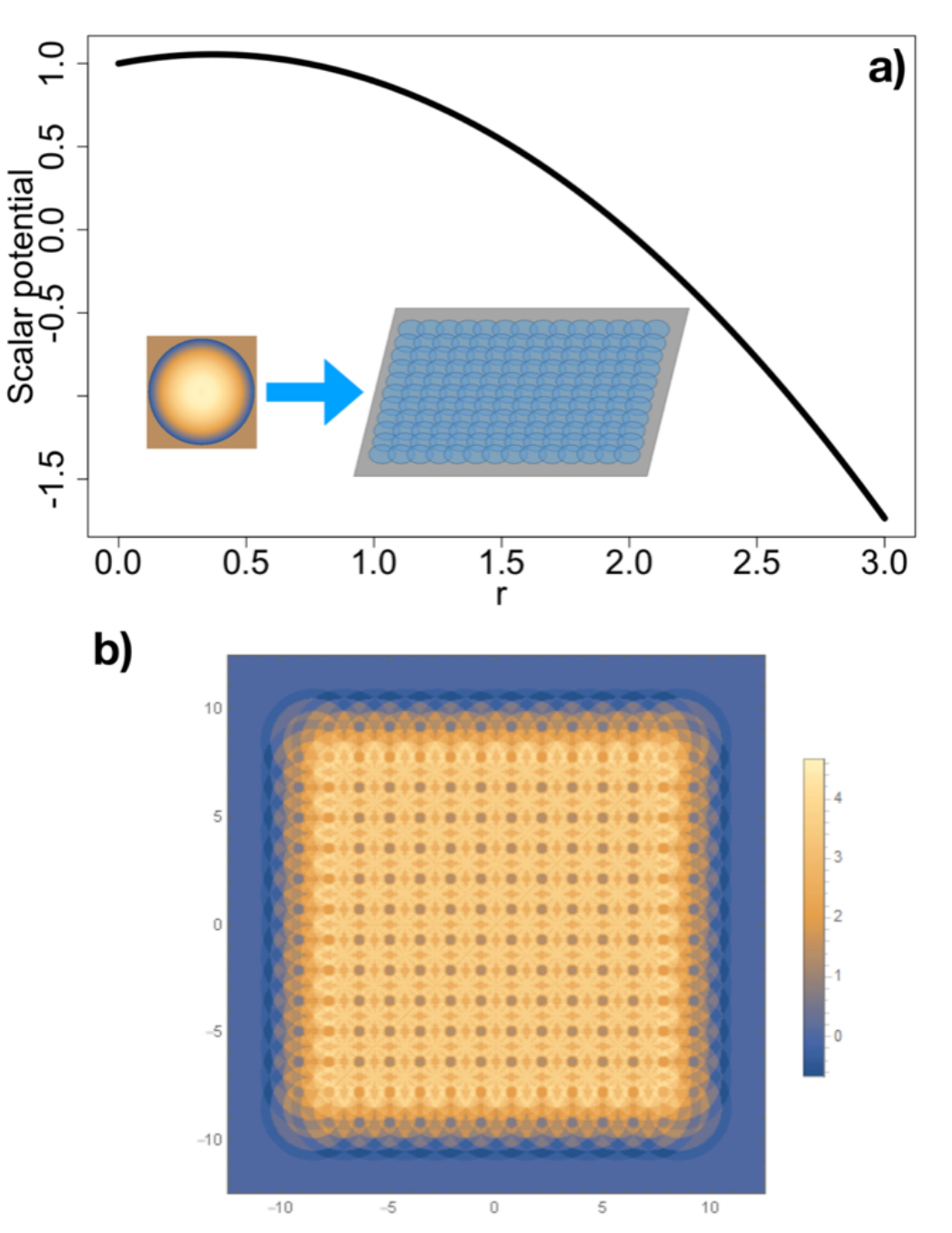}
\caption{a). The radial profile of the one-body potential approximating Eq.(\ref{mr}), when it is $\frac{3}{2}V_1^{\text{2bdy}}+\frac{1}{2}V_3^{\text{2bdy}}+\frac{2}{3}V_3^{\text{3bdy}}$. The inset shows the potential profile in the two-dimensional real space, and the schematics of the von Neumann lattice. b). The potential contour of the von Neumann lattice of the local potentials in a). This potential profile can lead to much larger incompressibility gap and ground state/quasihole state overlap to the model states, as compared to the Coulomb interaction in the second Landau level\cite{supp}}
\label{fig1}
\end{figure}

Some comments are in order here. For $n=1$ the vNL effectively adds a single two-body pseudopotential of $\hat V_1^{\text{2bdy}}$ for any positive $\lambda_0,\lambda_1$. This allows us to tune a single pseudopotential in an experimental FQH system, leaving the rest of the pseudopotentials from electron-electron interaction intact. For $n=2$ in Eq.(\ref{mr}), there are three tuning parameters $\lambda_0,\lambda_1,\lambda_2$. While we cannot obtain a pure $\hat V_3^{\text{3bdy}}$, we can nevertheless obtain the optimal two-body interaction\cite{peterson3} with $\left(2\lambda_{0,1,2}-\frac{3}{4}\lambda_{1,2}\right)/\left(\frac{3}{4}\lambda_{1,2}\right)=3$, and further enhanced with a single positive three-body interaction $\hat V_{3}^{\text{3bdy}}$. For example with $\lambda_{0,1,2}=1,\lambda_{1,2}=\frac{2}{3}$, Eq.(\ref{mr}) is effectively $\frac{3}{2}V_1^{\text{2bdy}}+\frac{1}{2}V_3^{\text{2bdy}}+\frac{2}{3}V_3^{\text{3bdy}}$. This vNL could be used to stabilise the Pfaffian state even in the lowest LL (see Fig.(\ref{fig1}), and also see some numerical results in \cite{supp}). One can similarly tune the anharmonic confining potential for $n=3$ to maximise $\hat V_4^{\text{2bdy}}$, or to design more complicated local potentials (beyond one-body) to enhance the incompressibility gap of the Fibonacci state. 

{\it Incompressibility gap --}
The equivalence of the vNL local projectors and the model Hamiltonians for the Read-Rezayi series allows us to treat the FQH effect as a ``continuous" version of the AKLT model. The Hamiltonian of the latter is given by a lattice of overlapping projection operators for the neighbouring spin Hilbert space. Like the AKLT model, the FQH ground state is ``frustration-free", satisfying the projection operators everywhere. The special property of the projection operators $\hat P^2=\hat P$ leads to rigorous proof of the ground state gap in the thermodynamic limit for the AKLT model\cite{aklt1,aklt2}. The lower bound of such gap can also be established from the gap of finite systems with open boundary condition\cite{aklt3,aklt4}. Thus the reformulation of the FQH model Hamiltonians can potentially lead to rigorous proof of the incompressibility gap in the thermodynamic limit as well, a long-standing problem of fundamental importance. 

While a rigorous proof is still not available, we discuss some interesting properties arising from this reformulation. Using Laughlin phase as an example, we write $\hat P_i= \hat T_{\vec X_i}|0,1\rangle\langle 0,1|\hat T^\dagger_{\vec X_i}$. Any \emph{translationally invariant} eigenstate $|\psi_\alpha\rangle$ with eigenvalue $\epsilon_\alpha$ gives
\begin{eqnarray}\label{gap}
\epsilon_\alpha=\frac{\mathcal A}{2\pi l_B^2}\langle\psi_\alpha|\hat P|\psi_\alpha\rangle
\end{eqnarray}
where $\mathcal A$ is the area of the Hall manifold, and $\hat P=|0,1\rangle\langle 0,1|$ is used because of translational invariance. Eq.(\ref{gap}) implies the energy gap of an excited state $|\psi_\alpha\rangle$ depends on the reduced density matrix of a circular droplet containing two single particle orbitals centered at the origin. More specifically, we can write $|\psi_\alpha\rangle=c_{\alpha 1}|\text{vac}\rangle|\tilde\psi_{\alpha 1}\rangle+c_{\alpha 2}|0\rangle|\tilde\psi_{\alpha 2}\rangle+c_{\alpha 3}|1\rangle|\tilde\psi_{\alpha 3}\rangle+c_{\alpha 4}|0,1\rangle|\tilde\psi_{\alpha 4}\rangle$, where $|\tilde\psi_{\alpha i}\rangle$ belongs to the Hilbert space outside of the two-flux droplet. For any state with finite energy $\epsilon_\alpha$ in the thermodynamic limit, we thus have $|c_{\alpha 4}|$ decaying at least as fast as $N_e^{-\frac{1}{2}}$. The necessary and sufficient condition for $\epsilon_\alpha$ to be finite and non-zero is thus:
\begin{eqnarray}\label{condition}
\lim_{N_e\rightarrow\infty}|c_{\alpha 4}|^2\sim \mathcal A^{-1}
\end{eqnarray}
Eq.(\ref{condition}) implies if there is a non-vanishing probability of finding two electrons in a circular drop containing two magnetic fluxes anywhere in the Hall fluid, the state has to be gapped. The excitation gap here refers to only translationally invariant excitations that have the same quantum numbers as the ground state. The ground state is incompressible if it is the only state where the LEC condition\cite{yang3} applies. Such uniqueness has been established numerically, which is another evidence apart from the finite size scaling of the energy gap itself.

{\it Summary and outlook --}
We show by forming a von Neumann lattice of local potentials in the Hall manifold, the effective Hamiltonian does not break translational symmetry in the Hilbert space of a single LL. Instead, it is analytically equivalent to short range projection Hamiltonians, representing two- or few-body pseudopotential interactions. In general a periodic potential in a single LL splits the degeneracy of the single particle orbitals and forms sub-bands\cite{cf,weimann,jain,halperin}. These are cases where the lattice spacing is much larger than the magnetic length. Nontrivial physics arises in the form of Hofstadter states or quantum anomalous Hall insulator\cite{thouless,hofstadter,macdonald,junren}. Interestingly, here with the vNL (i.e. lattice spacing $\sim\sqrt{2\pi}l_B$) there is effective translational invariance in a single LL, and the resulting Hamiltonians are equivalent to pure electron-electron interactions.

Experimentally, if it is technically possible to impose closely packed local potentials on the Hall bar and to accurately tune the potential profiles, the results in this work presents the exciting possibility of tuning individual PP (including three- or four-body PPs, etc), as well as the robust realisation of coveted non-Abelian FQH phases (e.g. the MR and the Fibonacci states). The spacing between local potentials needs to be on the order of $\sim 30 nm$ with $B=5\text{T}$, and $\sim 20 nm$ with $B=10\text{T}$. This is technically feasible using the formation of antidot arrays with the current e-beam processing technology. We would also like the local potentials to be overlapping: the range of the potential is larger than the potential spacing. It can be achieved if a spacer with proper thickness is synthesised between the antidot array and the Hall manifold. The accurate tuning of the local potentials could be more feasible in the cold atom systems, with more flexibility in tuning the entire photonic lattice\cite{marco,kimble}. The local projection approach may also be applicable to FQH like physics in spin systems\cite{nielsen}, leading to new platforms for non-Abelian topological states.

The construction of Eq.(\ref{h}) is analogous to the projection operators in quantum magnets, e.g. the AKLT model, where translationally invariant Hamiltonians are achieved by projection operators on every pair of neighbouring spins. In both cases, while the quantum Hamiltonian cannot impose local truncation of Hilbert space exactly (for spin chains, each spin is shared by different projection operators, analogous to non-orthogonality of coherent states in vNL), the ground states of such Hamiltonians nevertheless do satisfy local truncation of the Hilbert space everywhere and are thus free of frustration. The gapless edge states in FQH are also analogous to the ground state degeneracy of the open boundary AKLT chain. The reformulation of the model interaction Hamiltonian as local real-space operators may lead to analytical proofs of long-standing problems about the incompressibility of such Hamiltonians. Unlike the AKLT model, however, each projection operator in the vNL acts on the same Hilbert space no matter how far away they are separated. One thus cannot directly apply Knabe's argument\cite{aklt3} in determining the lower bound of the excitation gap. However the local Hilbert space shared by two operators decay exponentially with their separation. A more detailed analysis on the spectral gap of the vNL Hamiltonian will be presented elsewhere.

\begin{acknowledgments}
{\sl Acknowledgements.} This work is supported by the NTU grant for Nanyang Assistant Professorship and the National Research Foundation, Singapore under the NRF fellowship award (NRF-NRFF12-2020-005).
\end{acknowledgments}



\clearpage

\renewcommand{\thefigure}{S\arabic{figure}}
\renewcommand{\theequation}{S\arabic{equation}}
\renewcommand{\thepage}{S\arabic{page}}
\setcounter{figure}{0}
\setcounter{page}{1}

\onecolumngrid
\begin{center}
\textbf{\large Supplementary Online Materials for ``Fractional quantum Hall effect from frustration-free Hamiltonians"}
\end{center}
\setcounter{equation}{0}
\setcounter{figure}{0}
\setcounter{table}{0}
\setcounter{page}{1}
\makeatletter
\renewcommand{\theequation}{S\arabic{equation}}
\renewcommand{\thefigure}{S\arabic{figure}}
\renewcommand{\bibnumfmt}[1]{[S#1]}
\renewcommand{\citenumfont}[1]{S#1}
In this supplementary material, we give more technical details on how the effective Hamiltonians can be constructed from the von Neumann lattice (vNL) of projection operators. On an infinite plane such Hamiltonians are in the following general form:
\begin{eqnarray}\label{heff}
\mathcal H=\int\frac{d^2r}{2\pi l_B^2}|\psi(\vec X)\rangle\langle\psi(\vec X)|
\end{eqnarray} 
with $|\psi(\vec X)\rangle=\hat T_{\vec X}|\psi(\vec 0)\rangle$, where $\hat T_{\vec X}$ is the magnetic translation operator acting on all electrons in $|\psi(\vec 0)\rangle$, which is a Slater determinant state centered at the origin as described in the main text. In this case Eq.(\ref{heff}) is an $n_e-$body interaction if $|\psi(\vec 0)\rangle$ contains $n_e$ electrons, and in the second quantised for we have
\begin{eqnarray}
&&\mathcal H=\sum_{a_1,\cdots, a_{n_e},a'_1,\cdots,a_{n_e}}V_{a_1,\cdots,a_{n_e}}^{a'_1,\cdots,a'_{n_e}}c_{a'_1}^\dagger\cdots c_{a'_{n_e}}^\dagger c_{a_1}\cdots c_{a_{n_e}}\\
&&V_{a_1,\cdots,a_{n_e}}^{a'_1,\cdots,a'_{n_e}}=\langle a'_1,\cdots,a'_{n_e}|\mathcal H|a_1,\cdots,a_{n_e}\rangle\label{melement}
\end{eqnarray}
Here $0\le a_i,a'_i$ are non-negative integers. To compute the matrix elements in Eq.(\ref{melement}), let us write $|\psi(\vec 0)\rangle=|k_1,k_2,\cdots, k_{n_e}\rangle=c^\dagger_{k_1}c^\dagger_{k_2}\cdots c^\dagger_{k_{n_e}}|\text{vac}\rangle$ and we have the following
\begin{eqnarray}
V_{a_1,\cdots,a_{n_e}}^{a'_1,\cdots,a'_{n_e}}&=&\langle a'_1,\cdots,a'_{n_e}|\mathcal H|a_1,\cdots,a_{n_e}\rangle\nonumber\\
&=&\int\frac{d^2r}{2\pi l_B^2}\langle a'_1,\cdots,a'_{n_e}|\hat T_{\vec X}|k_1,k_2,\cdots, k_{n_e}\rangle\langle k_1,k_2,\cdots, k_{n_e}|\hat T^\dagger_{\vec X}|a_1,\cdots,a_{n_e}\rangle\label{integration}
\end{eqnarray}
It is easy to show that (without loss of generality we take $a_i\ge k_j$)
\begin{eqnarray}
\langle a_i|\hat T_{\vec X}|k_j\rangle=e^{-\frac{1}{4}|\vec X|^2}\sqrt{\frac{k_j!}{a_i!}}\left(\frac{i|\vec X|}{\sqrt 2}\right)^{a_i-k_j}e^{-i\left(a_i-k_j\right)\theta_{\vec X}}L_{k_j}^{a_i-k_j}\left(\frac{|\vec X|^2}{2}\right)
\end{eqnarray}
where $L_n^\alpha\left(x\right)$ is the generalised Laguerre polynomial, and $\theta_{\vec X}$ is the angle of $\vec X$. Thus the integration in Eq.(\ref{integration}) can be carried out explicitly by noting that $r^a=\epsilon^{ab}X_b$. One can thus check analytically the equivalence of Eq.(\ref{heff}) with the conventional pseudopotential model Hamiltonians. 

As an example, we take $|\psi(\vec 0)\rangle=|0,3\rangle$, so we are dealing with a two-body interaction. The relevant matrix elements are thus
\begin{eqnarray}
&&\hat{\mathcal H}^{\left(0,3\right)}=\sum_{a_1,a_2,a'_1,a'_2}V_{a_1,a_2}^{a'_1,a'_2}c_{a'_1}^\dagger c_{a'_{2}}^\dagger c_{a_1}c_{a_{2}},\quad V_{a_1,a_2}^{a'_1,a'_2}=\int\frac{d^2X}{2\pi l_B^2}\langle a'_1,a'_2|\hat T_{\vec X}|0,3\rangle\langle 0,3|\hat T^\dagger_{\vec X}|a_1,a_2\rangle\label{integration}
\end{eqnarray}
We can thus write $\hat{\mathcal H}^{\left(0,3\right)}=\lambda_1\hat V_1^{\text{2bdy}}+\lambda_2\hat V_3^{\text{2bdy}}$ with
\begin{eqnarray}
&&\lambda_1=\langle 0,1|\hat{\mathcal H}^{\left(0,3\right)}|0,1\rangle=\frac{3}{4}\\
&&\lambda_2=\left(\frac{1}{2}\langle 0,3|-\frac{\sqrt 3}{2}\langle 1,2|\right)\hat{\mathcal H}^{\left(0,3\right)}\left(\frac{1}{2}|0,3\rangle-\frac{\sqrt 3}{2}|1,2\rangle\right)=\frac{1}{4}
\end{eqnarray}
where $|0,1\rangle$ is the state of two particles with relative angular momentum $1$, while $\frac{1}{2}|0,3\rangle-\frac{\sqrt 3}{2}|1,2\rangle$ is the state of two particles with relative angular momentum $3$.  All other matrix elements relevant to pseudopotentials are zero. This allows us to write down the following exact relationship:
\begin{eqnarray}
\hat{\mathcal H}^{\left(0,3\right)}=\frac{3}{4}\hat V_1^{\text{2bdy}}+\frac{1}{4}\hat V_3^{\text{2bdy}}
\end{eqnarray}
For general projection operator with $|\psi\left(0,0\right)\rangle=|k_1,\cdots,k_{n_e}\rangle$, the expansion of the vNL effective Hamiltonian into a linear combination of pseudopotentials can be computed similarly. We will not repeat the detailed computation here, but to list a number of results that are used in the main text as follows:
\begin{eqnarray}
&&\hat{\mathcal H}^{\left(0,1\right)}=\hat V_1^{\text{2bdy}}\\
&&\hat{\mathcal H}^{\left(0,2\right)}=\hat V_1^{\text{2bdy}}\\
&&\hat{\mathcal H}^{\left(0,3\right)}=\frac{3}{4}\hat V_1^{\text{2bdy}}+\frac{1}{4}\hat V_3^{\text{2bdy}}\\
&&\hat{\mathcal H}^{\left(1,2\right)}=\frac{1}{4}\hat V_1^{\text{2bdy}}+\frac{3}{4}\hat V_3^{\text{2bdy}}\\
&&\hat{\mathcal H}^{\left(1,3\right)}=\frac{1}{2}\hat V_1^{\text{2bdy}}+\frac{1}{2}\hat V_3^{\text{2bdy}}\\
&&\hat{\mathcal H}^{\left(2,3\right)}=\frac{1}{8}\hat V_1^{\text{2bdy}}+\frac{1}{4}\hat V_3^{\text{2bdy}}+\frac{5}{8}\hat V_5^{\text{2bdy}}\\
&&\hat{\mathcal H}^{\left(0,1,2\right)}=\hat V_3^{\text{3bdy}}\\
&&\hat{\mathcal H}^{\left(0,1,3\right)}=\hat V_3^{\text{3bdy}}\\
&&\hat{\mathcal H}^{\left(0,2,3\right)}=\frac{1}{3}\hat V_3^{\text{3bdy}}+\frac{2}{3}\hat V_5^{\text{3bdy}}\\
&&\hat{\mathcal H}^{\left(1,2,3\right)}=\frac{1}{27}\hat V_3^{\text{3bdy}}+\frac{2}{9}\hat V_5^{\text{3bdy}}+\frac{20}{27}\hat V_6^{\text{3bdy}}\\
&&\mathcal H^{\left(0,1,2,3\right)}=\hat V_6^{\text{4bdy}}
\end{eqnarray}

Various vNL effective Hamiltonians with local anharmonic well as mentioned in the main text are obtained as linear combinations of the expressions listed above. In Fig.(1) of the main text, we show the real-space potential profile of the vNL given in Eq.(8) of the main text, by taking $\lambda_{0,1,2}=1,\lambda_{1,2}=\frac{2}{3}$. A quick numerical analysis with $14$ electrons can show it is much superior in supporting the non-Abelian Moore Read phase as compared to the realistic $V_{\text{SLL}}$, the Coulomb interaction in the second Landau level. The incompressibility gap of $V_{\text{SLL}}$ is $0.011$(after normalising the $\hat V_1^{\text{2bdy}}$ component to unity), while that from Eq.(2) is $0.55$, an order of magnitude larger. This indicates a much more robust Hall plateau. In terms of the wavefunction overlaps to the model ground state and four quasihole states containing two quasiholes (insertion of one magnetic flux), the eigenstates of $V_{\text{SLL}}$ have the overlap of $0.693$ for the ground state, and $0.61, 0.35,0.62, 0.52$ for the four quasihole states. In contrast, the eigenstates from Eq.(2) have overlap of $0.998$ for the ground state, and $0.98, 0.99, 0.99, 0.98$ for the four quasihole states. The good overlap with the quasihole states is also very important for the non-Abelian statistics from the braiding of the quasiholes. Similar results can also be obtained for the Fibonacci states from Eq.(9) of the main text, though we cannot go to very large system sizes with four-body interaction in numerical computations.  
\end{document}